\documentstyle[sprocl]{article}

\def\be{\begin{equation}}
\def\ee{\end{equation}}
\def\bea{\begin{eqnarray}}
\def\eea{\end{eqnarray}}

\begin{document}

\title{Thermodynamics of the Casimir Effect\\
        - Asymptotic Considerations -}

\author{H. Mitter, D. Robaschik}

\address{Institut f\"ur Theoretische Physik der 
Karl-Franzens-Universit\"at Graz, Universit\"atsplatz 5 \\
A-8020 Graz, Austria}      


\maketitle\abstracts{ 
We study the Casimir effect with different temperatures
between the plates ($T$) resp. outside of them ($T'$).
If we consider the inner system as the 
black body radiation for a special geometry, then contrary
to common belief the temperature approaches a constant value 
for vanishing volume during isentropic processes.
This means:  the reduction of the degrees of 
freedom  can not be compensated by a concentration of the energy
during an  adiabatic contraction of the two-plate system. 
Looking at the Casimir pressure, we find
one unstable equilibrium point for isothermal
processes  with $T > T'$. 
For isentropic processes there is 
additionally one stable equilibrium point for larger values
of the distances between the two plates.}

\section{Introduction}
\noindent 
The Casimir effect \cite{CAS} is one of the fundamental effects 
of Quantum Field Theory. It tests the importance of the zero point
energy. In principle, one considers two conducting infinitely
extended parallel plates at the positions $x_3=0$ and $x_3= a$.
These conducting plates change the vacuum energy of Quantum 
Electrodynamics in such a way that a measurable attractive     
force between both plates can be observed 
\cite{EXP}. This situation does not essentially change, if  
a nonvanishing temperature \cite{MF} is taken into account.
The thermodynamics
of the Casimir effect \cite{BML} \cite{GREIN} 
and  related problems 
\cite{BARTO} 
is well investigated.\\
Here we shall treat the different regions separately.
We assume a temperature $T$ for the space between
the plates and a  temperature $ T' $ for the space outside the plates. 
Thereby we consider the right plate at $ x_3=a $ as movable, so
that different thermodynamic processes such as isothermal or 
isentropic motions, can be studied.
At first we investigate the thermodynamics of the space between
the two plates by setting $T'=0$. This can be viewed as the black 
body radiation
(BBR) for a special geometry. The surprising effect is, that for
vanishing distance ($a\rightarrow 0$) in isentropic processes
the temperature approaches a finite value, which is completely
determined by the fixed entropy. This is in contrast to the 
expected behaviou of the standard BBR, if the
known expression derived for a large volume is extrapolated
to a small volume.
For large values of $a$ the BBR takes the standard form.
As a next topic we consider the Casimir pressure assuming   
that the two physical regions, i.e. the spaces between and 
outside the two plates possess different temperatures. 
Depending on  the choices of $T$ and $T'$ 
a different 
physical behaviour is possible.
For $T'<T$ the external pressure is reduced in comparison with the
standard case $T'=T$. Therefore we expect the existence of
an equilibrium point, where the pure Casimir
attraction  ($T=0$ effect ) and the 
differences of the radiation pressures compensate each other. 
This point is unstable, so that for isothermal processes 
the movable plate moves either to 
$a\rightarrow 0$ or to $a \rightarrow \infty$. However, an 
isentropic motion reduces the internal radiation pressure
for growing distances, so that in this case
there is an additional stable equilibrium point.

\section{Thermodynamic Functions}
The thermodynamic functions are already determined by different
methods \cite{MF} \cite{BML}. We recalculate them by 
statistical mechanics including the zero-point energy and cast
it in a simpler form which can be studied in detail \cite{MR}. 
For technical reasons the 
system is embedded in a large cube (side L). As space between the 
plates 
we consider the volume $L^2a$, the region outside is given by 
$L^2(L-a)$. All extensive thermodynamic functions are defined per area.
\\
Free energy $\phi = F/L^2$:
\begin{eqnarray}
\label{1} 
\phi_{int} &=& [\frac{\hbar c \pi^2}{a^4}(-\frac{1}{720} + g(v))
              +\frac{3\hbar c}{\pi^2} \frac{1}{\lambda^4} ]a ,\\
\label{2}
\phi_{ext} &=& [\frac{3\hbar c}{\pi^2} \frac{1}{\lambda^4} 
                -\frac{\hbar c \pi^6}{45}(\frac{v'}{a})^4](L-a).
\end{eqnarray}
Energy $e = E/L^2$:
\begin{eqnarray*}
e_{int} &=& [\frac{\hbar c \pi^2}{a^4}(-\frac{1}{720} + g(v)
              -v \partial_v g(v))
              +\frac{3\hbar c}{\pi^2} \frac{1}{\lambda^4} ]a ,\\
e_{ext} &=& [\frac{3\hbar c}{\pi^2} \frac{1}{\lambda^4} 
              +\frac{3 \hbar c \pi^6}{45}(\frac{v'}{a})^4](L-a).
\end{eqnarray*}
Pressure:
\begin{eqnarray}
\label{3} 
p_{int} &=& [\frac{\hbar c \pi^2}{a^4}(-\frac{1}{240} + 3g(v)
              -v\partial_v g(v))
              -\frac{3\hbar c}{\pi^2} \frac{1}{\lambda^4} ],\\
\label{4}
p_{ext} &=& [\frac{3\hbar c}{\pi^2} \frac{1}{\lambda^4} 
                -\frac{\hbar c \pi^6}{45}(\frac{v'}{a})^4].
\end{eqnarray}
Entropy $\sigma = S/(k L^2)$:
\begin{eqnarray}
\label{5} 
\sigma_{int} = -\frac{ \pi}{a^3} \partial_v g(v) a;\;\;\,
\sigma_{ext} =  \frac{4\pi^5}{45} (\frac{v'}{a})^3 (L-a), 
\end{eqnarray}
$\lambda$ regularizes  ($\lambda \rightarrow 0 $)
the contributions from the zero-point 
energy. The thermodynamics is governed by the function $g(v)$.
We list two equivalent expressions:
\begin{eqnarray}
\label{6} 
g(v) = -v^3[\frac{1}{2}\zeta (3) + k(\frac{1}{v})] =
    \frac{1}{720} -\frac{\pi^4}{45}v^4 - 
   \frac{v}{4\pi^2}[\frac{1}{2} \zeta(3) + k(4\pi^2 v)].
\end{eqnarray}
The function $k(x)$ is given by
\begin{eqnarray}
\label{7} 
k(x) = (1- x\partial_x)\sum_{n=1}^{\infty}
      \frac{1}{n^3}\frac{1}{exp(nx) - 1}.
\end{eqnarray}
It is strongly damped for large arguments. $v$ is the known variable
$v = a  T k/(\hbar \pi c)$, the variable $v'$ contains 
the temperature $T'$ instead of $T$. 

\section{Black Body Radiation}
\noindent 
As a first topic we consider the space between the two plates as
a generalization of the usual black body radiation (BBR) for a 
special geometry $L \times L \times a $. Contrary to the 
standard treatment we include here both, the internal and 
external the zero point energy.
Thereby parameter-dependent divergent contributions compensate 
each other, whereas the physically irrelevant
term $ ~ L/{\lambda^4}$ can be omitted \cite{MR}.
If we approximate the function $g$ for large $v$ 
by $g \simeq {1}/{720} - (\pi^4/45) v^4 
- \zeta(3)/(8\pi^4) v $, we obtain
\begin{eqnarray}
\label{8}
\phi_{as} &=& \frac{\pi^2 \hbar c}{a^3}[-\frac{\pi^4}{45}v^4
            -\frac{\zeta(3) }{8\pi^2} v],\;\;\;
\sigma_{as} = \frac{\pi}{a^2} [\frac{4\pi^4}{45}v^3
            +\frac{\zeta(3) }{8\pi^2} ],\\
\label{9}
p_{as}&=& \frac{\pi^2 \hbar c}{a^4} [\frac{\pi^4}{45}v^4
            -\frac{\zeta(3) }{8\pi^2} v],\;\;\;\
e_{as}= \frac{\pi^2 \hbar c}{a^3} \frac{3\pi^4}{45}v^4.            
\end{eqnarray}
These expressions contain the large-volume
 contributions 
corresponding to the standard BBR (first term) 
and corrections.
In the other limit of small $v$,
we have to use $g(v) = - v^3 \zeta(3)/2 $ and get
\begin{eqnarray}
\label{10}
\phi_{o} &=& \frac{\pi^2 \hbar c}{a^3}[-\frac{1}{720}
            -\frac{\zeta(3) }{2} v^3],\;\;\;
\sigma_{o} = \frac{\pi}{a^2} 
             \frac{3 \zeta(3) }{2} v^2, \\
\label{11}
p_{o}&=& \frac{\pi^2 \hbar c}{a^4} [-\frac{1}{240}],\;\;\;\
e_{o}= \frac{\pi^2 \hbar c}{a^3} [-\frac{1}{720}
            +\zeta(3)  v^3].
\end{eqnarray}
In this case the contributions of the zero point energy
dominate. It is known that nondegenerate vacuum states
do not  contribute to the entropy, which indeed vanishes at 
$T=0$.\\
Let us now consider isentropic processes. This means 
that we fix the values of the entropy for the internal
region (\ref{5})  during the complete process. 
Technically we
express this fixed value according to
the value of the variable $v$ either through the approximation 
(\ref{8}) or (\ref{10}).
 Large distances 
and/or
high temperatures  lead to large values of $v$ so we have to
use $\sigma_{as}$. Constant  entropy means
\begin{eqnarray}
\label{12} 
\sigma = {\rm const.} =
\sigma_{as} =  \frac{4\pi^2k^3}{45(\hbar c)^3} a T^3
                        +\frac{\zeta(3)}{8\pi} \frac{1}{a^2 }.
\end{eqnarray}
Asymptotically this is the standard relation
BBR  $S = L^2 \sigma_{as}
= {\rm const.} \times V T^3 $, here valid for large $T $ and 
$V$. If we now consider smaller values of $a$, then, because of
eq.(\ref{5}), 
also $-\partial_v g(v)$
takes  smaller values. It is possible to prove \cite{MR}
the inequalities
$ g <0 $, $ \partial_v g(v) <0 $ and $ (\partial_v)^2 g(v) <0 $.
This monotonic behaviour of $\partial_v g(v)$ leads to the 
conclusion that also the corresponding values of $v$ become 
smaller. Consequently,  we have to apply the
other represention (\ref{10}) for small $v$
and obtain 
\begin{eqnarray}
\label{13} 
\sigma =\sigma_{as}=\sigma_{o} 
 =\frac{k^2}{\hbar^2 c^2 \pi} 
            \frac{3 \zeta(3) }{2} T^2. 
\end{eqnarray}
This means that for $ a \rightarrow 0 $  the temperature
does not tend to infinity, but approaches the finite value
\begin{eqnarray}
\label{14} 
T =  \left(\sigma \,\, 2 \hbar^2 c^2 \pi/(3 \zeta(3) k^2)
     \right)^{1/2}.
\end{eqnarray}
This is in contrast to the expectation: if we apply the 
standard expression of  BBR, fixed entropy implies 
$VT^3 = {\rm const.} $, so that the temperature tends to 
infinity for vanishing volume. 
However this standard expression for BBR, 
derived for a continuous frequency spectrum, is not valid
for small distances. The reduction of the degrees of freedom, 
i.e. the transition from a continuous frequency spectrum to 
a discrete spectrum, is the reason for our result.

\section{Equilibrium Points of the Casimir Pressure}
\noindent 
The Casimir pressure results from the contributions of 
the internal and the external regions acting on the right  
movable plate.  
\begin{eqnarray}
\label{15} 
P(a,T,T') = P_{ext}(T') + P_{int}(a,T) 
          = \frac{\pi^2 \hbar c}{a^4}p(v) 
            +\frac{\pi^2 k^4}{45 (\hbar c)^3}(T^4 - {T'}^4), 
\end{eqnarray}
where  
\begin{eqnarray}
p(v) = -\frac{1}{4\pi^2} v[\zeta(3) +(2 - v\partial_v)k(4\pi^2v)]
      =-\frac{1}{240} +3g(v) - v\partial_{v}g(v) 
         -\frac{\pi^4}{45} v^4.\nonumber
\end{eqnarray}
Usually one considers the case $T=T'$, so that the Casimir
pressure is prescribed by $p(v)$ alone. 
It is known, that 
$P(a,T,T'=T)$ is a negative but monotonically rising function 
from $-\infty$ (for $ a\rightarrow 0  $) to $ \; 0\; $ (for 
$a\rightarrow \infty $). 
It is clear, that the addition of a positive pressure
$ \frac{\pi^2 k^4}{(\hbar c)^3}(T^4 - {T'}^4) $ for $T>T'$
stops the Casimir attraction at a finite value of $ v$.
The question is whether this equilibrium point may be
stable or not? The answer 
follows from the monotonically rising behaviour of the standard
Casimir pressure. 
\begin{eqnarray}
\label{16}
\frac{d}{da}P(a,T,T') =\frac{d}{da}P(a,T,T'=T) >0.
\end{eqnarray}
Consequently this equilibrium point is unstable 
(see also \cite{MR}). \\
Next we consider the space between the two plates not for fixed
temperature but as a  thermodynamically closed system with 
fixed entropy. In the
external region we assume again  a fixed 
temperature $T'$.
To solve this problem in principle, it is sufficient to discuss
our system for large $v$ (as large $v$ we  mean such 
values of $v$ for which the asymptotic approximations (\ref{8}),
(\ref{9}) are valid; this region starts at  
 $ v> 0.2 $ ). 
Using our asymptotic 
formulae (\ref{8}),(\ref{9}) we write the Casimir pressure 
as
\begin{eqnarray}
\label{17}
P(a,v,T') = \frac{\pi^2 \hbar c}{a^4}[\frac{\pi^4}{45}v^4
             -\frac{\zeta(3)}{4\pi^2 }v
              - \frac{\pi^4}{45}{v'}^4 ],                        
\end{eqnarray}
with $v' = aT' k/(\hbar c \pi)$ where $v$ has to be determined
from the condition $ \sigma_{as}=\sigma = {\rm const.} $ or
\begin{eqnarray}
\label{18}
 \pi v^3 = [ a^2 \sigma - \zeta(3)/(8\pi^2)] 45/(4\pi^4).
\end{eqnarray}
Then we may write
\begin{eqnarray}
\label{19}
P(a,v,T') = \frac{\pi^2 \hbar c}{a^4}[\frac{\sigma a^2}
             {4\pi}  -\frac{9\zeta(3)}{32\pi^2 }]
    \{\frac{45}{4\pi^4}(\frac{\sigma a^2}{4\pi} 
   - \frac{\zeta(3)}{8\pi^2}) \}^{3/2}
          -\frac{\pi^2 \hbar c}{a^4} \frac{\pi^4}{45}{v'}^4.                        
\end{eqnarray}
At first we consider the case $T'=0$. 
We look for the possible 
equilibrium points $P(a,v,T'=0) =0$. The result is 
$ v^3 = 45\zeta(3)/(4 \pi^6)$. This corresponds to $v=0.24$. 
For this value of $v$ the used approximation is not very good, but
acceptable. 
A complete numerical estimate  \cite{MR} gives the same value.
Now we express the temperature $ T$ included in $v$ with the
help of the equation for isentropic motions 
(\ref{18}) and obtain
$ a^2 = 9\zeta(3)/(8 \pi \sigma)$. 
The instabiliy of this point can be directly seen by looking
at 
\begin{eqnarray}
\label{20}
\frac{d}{da}P(a,T,T'=0) &=& - 4 P(a,T,T'=0) 
     + \frac{\pi^2 \hbar c}{a^4}
[\frac{4 \pi^4}{45}v^3
             -\frac{\zeta(3)}{8\pi^2 }]
 (\frac{dv}{da })_{\sigma} |_{P=0}  \nonumber\\
&= & \frac{\pi^2 \hbar c}{a^4}\frac{3\zeta(3)}{4\pi^2}
             (\frac{dv}{da })_{\sigma}.
\end{eqnarray}
It is intuitively clear that
$(\frac{dv}{da })_{\sigma}$ is positive; an explicit proof
is given in \cite{MR}.
So it is clear,  that this point is unstable
as in the isothermal case. If we consider, in 
eq.(\ref{17}), the variable $v= aTk/(\hbar c \pi) $ at
fixed $T$, there is no further equilibrium point.
This result for isothermal processes  is, however, not
valid for isentropic processes. In this case we obtain 
according to eq.(\ref{19}) a second trivial equilibrium  point 
at $a \rightarrow \infty $ for vanishing external temperature 
($v'=0$).  
Between both zeroes we have one maximum. So we conlude: 
For isentropic processes there must be
two equilibrium points; the left one is unstable, the
right one at $  a \rightarrow \infty $  corresponds to a
vanishing derivative. If we now
add a not too high external pressure with the help of an external
temperature $T'$, then this second equilibrium point 
- present for isentropic processes - becomes stable. 
So, in principle  we may observe oscillations 
at the position of the second equilibrium point.
\section*{Acknowledgments}
We would like to thank C. B. Lang and N. Pucker
for their constant support and K. Scharnhorst, G. Barton and
P. Kocevar for  discussions on the present topic.


\section*{References}
\begin{thebibliography}{99}

\bibitem{CAS}
H. B. G. Casimir, {\em Proc. Kon. Ned. Akad. Wetenschap.} 
{\bf 51}  793 (1948).
\bibitem{EXP}
M. J. Sparnay, {\em Physica} {\bf 24},751 (1958);
S. K. Lamoreaux, {\em Phys. Rev. Lett.} {\bf 78}  5 (1997);
U. Mohideen, A. Roy, Preprint {\bf physics/9805038}. 
%
\bibitem{MF}
M. Fierz, {\em Helv. Phys. Acta} {\bf 33} , 855 (1960);
J. Mehra, {\em Physica} {\bf 37} 145 (1967).
%
\bibitem{BML}
L. S. Brown, and G. J. MacLay, {\em Phys. Rev}
 {\bf 184}  1272 (1969);
J. Schwinger, L.L. DeRaad, K.A. Milton,
{\em Annals of Physics (NY)}
{\bf 115}  1 (1978);
K. Scharnhorst, D. Robaschik, and E. Wieczorek, 
{\em Annalen d. Physik
(Leipzig)} {\bf 44}  351 (1987);  
D. Robaschik, K. Scharnhorst, and E. Wieczorek, 
{\em Annals of Physics (NY)}
{\bf 174}  401 (1987).
\bibitem{GREIN}
G. Plunien, B. M\"uller, and W. Greiner, 
{\em Phys. Reports} {\bf 134} (1986)
87 (1986);
G. Barton, and N. S. J. Fawcett, {\em Phys. Reports}
 {\bf 170}  1 (1988);
V. M. Mostepanenko, N. N. Trunov, {\em The Casimir Effect and its
Application, Oxford, 1997}.
\bibitem{BARTO}
G. Barton, J.Phys. A : {\em Math. Gen.} {\bf 24}  5533 (1991);
%
M. Revzen, R. Opher, and A. Mann, J. {\em Physics A : Math. Gen.}
 {\bf 30}
 7783 (1997), {\em Europhys. Lett.} {\bf 38}  245 (1997);
%
F. Ravndal, and D. Tollefsen {\em Phys. Rev.} {\bf 40 }  4191 (1989).
%
\bibitem{MR}
H. Mitter, D. Robaschik, to be published.

\end{thebibliography}
\end{document}